\documentclass[conference]{IEEEtran}

\usepackage{hyperref}
\hypersetup{
     colorlinks = true,
     citecolor = green,
     }

\ifCLASSINFOpdf
  \usepackage[pdftex]{graphicx}
\else
\fi
\usepackage{graphicx}
\usepackage{subcaption}
\usepackage{float}

\usepackage{xcolor}
\definecolor{cs-blue}{rgb}{0.035, 0.114, 0.235}
\definecolor{turq}{rgb}{0.365, 0.835, 0.851}
\definecolor{lightblue}{rgb}{0.8, 0.9, 1.0}

\usepackage{framed}



\usepackage{tcolorbox}

\tcbset{
  takeawaybox/.style={
    colback=turq!7!white,
    colframe=turq,
    coltitle=black,
    fonttitle=\bfseries\small,
    title=Key Takeaways:
  }
}

\usepackage{balance}

\usepackage{booktabs}
\usepackage{multirow}
\usepackage{amsmath}

\begin{document}

\title{Code Health in LLM-Based Test Generation:\\Effectiveness and Token Efficiency}


\author{
\IEEEauthorblockN{Freya Wirdemann}
\IEEEauthorblockA{\textit{Heidelberg University}\\
Heidelberg, Germany \\
is316@stud.uni-heidelberg.de}
\and
\IEEEauthorblockN{Markus Borg}
\IEEEauthorblockA{\textit{CodeScene and Lund University}\\
Malmö, Sweden \\
markus.borg@codescene.com}
\and
\IEEEauthorblockN{Nadim Hagatulah}
\IEEEauthorblockA{\textit{Lund University}\\
Lund, Sweden \\
nadim.hagatulah@cs.lth.se}
\and
\IEEEauthorblockN{Adam Tornhill}
\IEEEauthorblockA{\textit{CodeScene}\\
Malmö, Sweden \\
adam.tornhill@codescene.com}
}

%\author{\IEEEauthorblockN{Markus %Borg\IEEEauthorrefmark{1}, Amogha Borg\IEEEauthorrefmark{2}}
%\IEEEauthorblockA{\IEEEauthorrefmark{1}CodeScene, Malmö, Sweden and Lund University, Lund, Sweden, markus.borg@codescene.com}
%\IEEEauthorblockA{\IEEEauthorrefmark{2}CodeScene, Malmö, Sweden, amogha.udayakumar@codescene.com}}


\maketitle

\begin{abstract}
Coding agents powered by Large Language Models (LLMs) are now prominent in software engineering. Previous work has shown that AI tools perform better on high-quality source code that is easy to maintain. In this study, we investigate how the effectiveness of LLM-generated unit tests varies across maintainability levels measured by CodeScene's CodeHealth (CH). We assess test effectiveness using traditional coverage metrics and mutation score across Python, Java, and C++. Moreover, we study how code with different levels of CH translates into input tokens using common industrial tokenizers. Our results suggest that CH provides a weak but consistent signal of LLM-generated test effectiveness and is negatively correlated with input-token count. These findings provide further evidence for a relationship between maintainability and LLM-based software development. %These findings further emphasize the importance of maintaining high-quality code in the era of AI-assisted software development.
\end{abstract}


% \fi
%\IEEEpeerreviewmaketitle

\section{Introduction} \label{sec:intro}
%Software engineering is being disrupted by the widespread adoption of coding agents. Compared to previous industrial revolutions, the biggest difference in the AI era is the rapid speed of change. Research and practice are scrambling to learn what works and what does not, while simultaneously deploying the technology. Consequently, there is a strong need for empirical studies that can guide AI adoption in industry practice.
Industry observations reported by Thoughtworks indicate that AI tools, just like humans, perform better when operating on well-factored and modular code~\cite{thoughtworksTechnologyRadarOpinionated2025}. They refer to this as ``AI-friendly code design.'' Our recent work observed this phenomenon in a study of how code smells affect LLMs' ability to preserve semantic behavior when refactoring competitive programming solutions in Python~\cite{borgCodeMachinesNot2026}. While refactoring is an informative proxy for how LLMs process source code with varying levels of maintainability, it captures only a single dimension of software development.

In this study, we return to the question of how maintainability influences the performance of AI tools. We revisit the public dataset used in previous work and expand to roughly five times its size by adding C++ and Java solutions, each with varying numbers of code smells. Moreover, we shift the task from refactoring to test case generation, which we argue is another meaningful proxy for software development: we study how capable an LLM is at understanding existing code and generating new code from it. Moreover, we add a discussion of token counts, which have become a first-order cost concern, as illustrated by GitHub Copilot's shift to usage-based billing on June 1, 2026.

In line with the previous study, we characterize different maintainability levels using the CodeHealth\textsuperscript{TM} (CH) metric. CH has been validated as predictive of defects and
development effort in previous studies~\cite{tornhillCodeRedBusiness2022,borgIncreasingNotDiminishing2024}, performs well in benchmarking~\cite{borgGhostEchoesRevealed2024a}, and used operationally as a quality gate in CI and code review within the CodeScene product~\cite{martenssonMuchMoreTest2025,tverdalCombiningInsightsMultiple2025}. Based on this, we let the following research questions guide our study:

\begin{itemize}
    \item[RQ\textsubscript{1}] How is CodeHealth associated with the effectiveness of LLM-generated test cases?
    \item[RQ\textsubscript{2}] How does LLM input token usage vary across different levels of CodeHealth?
\end{itemize}

For RQ\textsubscript{1}, we operationalize test effectiveness using both standard code coverage metrics and mutation score~\cite{siami_namin_sufficient_2008}. Our findings in Sec.~\ref{sec:res} reveal a weak but consistent relationship between CH and the effectiveness of the generated test cases, and the results are more pronounced for mutation scores. For RQ\textsubscript{2}, we find that lower-CH code generally requires more input tokens, which persists after controlling for SLOC and is robust across contemporary tokenizer families. We discuss the practical implications and threats to validity in Sec.~\ref{sec:impl}.

\section{Background and Related Work} \label{sec:rw}
Test case generation has been a popular topic in test automation research~\cite{anandOrchestratedSurveyMethodologies2013}. Analogous to other research topics, the field has been disrupted by LLMs, and we focus this section accordingly. 

Jain \textit{et al.} developed the benchmark TestGenEval, based on SWEBench, to measure LLM-based test generation performance for Python~\cite{jainTestGenEvalRealWorld2025}. They evaluated models ranging from 7B to 405B, and reported the best results for OpenAI's GPT-4o: line coverage 35.2\% and mutation score 18.8\%, while smaller models generally performed worse.

We identified three relevant systematic literature reviews on LLMs and testing. Wang \textit{et al.} had the broadest scope and found that unit test generation was among the most common applications~\cite{wangSoftwareTestingLarge2024}. Chu \textit{et al.} focused on unit test generation and present a taxonomy based on 115 primary studies~\cite{chuLargeLanguageModels2025}. With a similar scope, Zhang \textit{et al.} presented a review based on 105 primary studies and report that coverage results vary greatly across studies~\cite{zhangLargeLanguageModels2025}. Both unit-testing reviews conclude that existing research has focused heavily on Python and Java.%, which motivates our inclusion of C++.

Ouédraogo \textit{et al.} present one of the most comprehensive studies of LLM-based unit test generation for Java~\cite{ouedraogoPromptEngineeringLLMs2026}. Their findings include that only 7.2\% of the generated tests compiled, with hallucinated symbols and incorrect API calls among the main causes of compilation failures. Moreover, they report median line coverage ranging from 3.57\% to 100\% across their three datasets, and explain this variation by differences in structural complexity and architectural diversity. %In this study, we investigate a related explanation by analyzing how code smells are associated with the effectiveness of LLM-generated tests.

%Another problem with LLM generated tests is that a significant amount is not compiling leading to lower test coverage and defect detection even though the inner test logic is correct~\cite{yangEvaluationLargeLanguage2024}. A solution here can be post-processing the tests.

%Even though larger LLMs show higher quality output across different tasks, the pass rate of generated tests is just as high as for smaller LLMs~\cite{shangLargeScaleEmpiricalStudy2025}. This indicates that using a larger LLM does not solve the problem of generating correct tests, which truly test the logic of the code.  
%Across different model sizes the fail rate of generated tests remains high, suggesting, that the pass rate of tests is not dependent of the model as much as other quality factors of the tests~\cite{chuLargeLanguageModels2025}

%Recently, iterative test case generation in agentic loops has been explored and the results are promising. 

%Dakhel et al. used feedback loops to refine the prompt based on surviving  mutants~\cite{dakhelEffectiveTestGeneration2024}. 
%Yoshimoto reports that AI agents show stronger test quality with, on average, higher statement and branch coverage gains than human-improved tests~\cite{yoshimotoTestingAIAgents2026}.
 
%Test generation remains one of the most challenging tasks for LLM-based agents across multiple programming languages, including Java, Python, and C++~\cite{sonwaneOmniCodeBenchmarkEvaluating2026}. The performance of different agents varies significantly, leading to substantial differences in test quality.

\section{Method} \label{sec:method}

\subsection{Dataset Creation} \label{sec:method-dataset}
Our starting point is the 5,000 Python files sampled from the CodeContests dataset~\cite{liCompetitionLevelCodeGeneration2022} analyzed in our previous refactoring study~\cite{borgCodeMachinesNot2026}. One limitation with that dataset is that it barely contains any examples with CH $< 8$. 
To remedy this limitation and allow us to study a broader range of CH scores, we complement the dataset by also sampling C++ and Java solutions, which contain more low-CH files.

We first calculated CH for all valid solutions. Following the stratified sampling strategy of the original study~\cite{borgCodeMachinesNot2026}, we sampled $2{,}500$ solutions with CH $\geq 9$ and $2{,}500$ with CH $< 9$ per language. We then added the lowest-scoring remaining files to improve coverage of the lower CH range, targeting approximately $10{,}000$ files per language. To encourage syntactic diversity, we applied CodeBLEU~\cite{renCodeBLEUMethodAutomatic2020} as a similarity filter in each sampling stage, accepting only candidates below a predefined similarity threshold. The final sets contained $9{,}795$ Java files and $9{,}825$ C++ files.

%We used the following weights: $n$-gram=$0.1$, weighted-$n$-gram=$0.4$, AST-match=$0.5$, and dataflow-match=$0$. For each candidate, CodeBLEU was computed against all already selected solutions. A candidate was only included if its maximum similarity to any selected solution was below $0.90$. 

\begin{figure}[t]
    \centering
    \begin{subfigure}[b]{\linewidth}
        \centering
        \includegraphics[width=\linewidth]{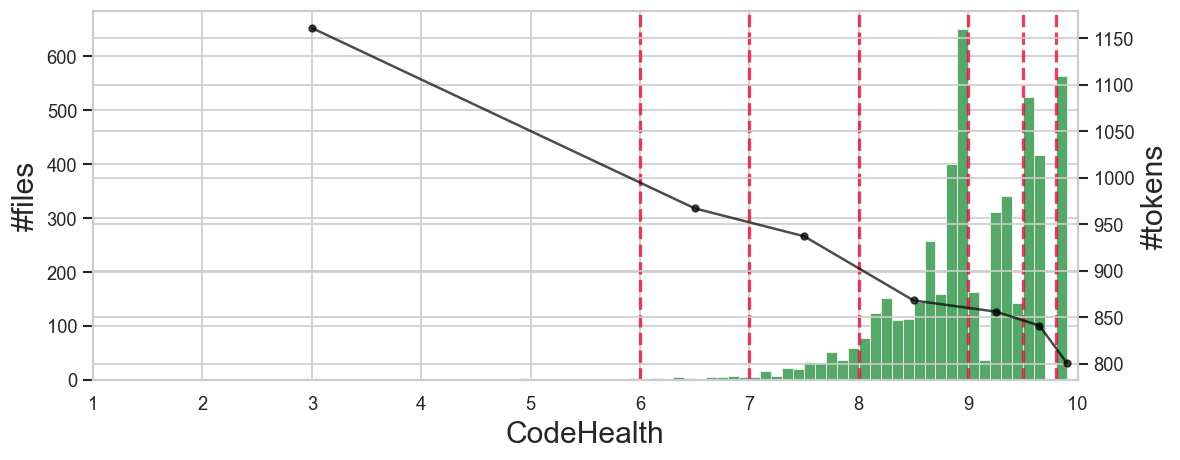}
        \caption{Python (n=5,000)}
        \label{fig:ch-dist-python}
    \end{subfigure}  
    \begin{subfigure}[b]{\linewidth}
        \centering
        \includegraphics[width=\linewidth]{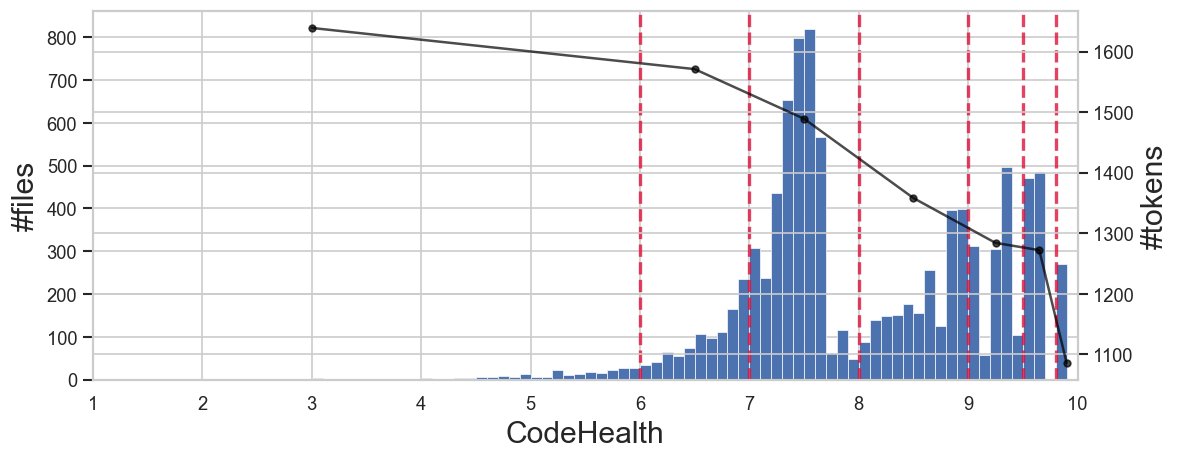}
        \caption{Java (n=9,795)}
        \label{fig:ch-dist-java}
    \end{subfigure}
    \vspace{0.5em}
    \begin{subfigure}[b]{\linewidth}
        \centering
        \includegraphics[width=\linewidth]{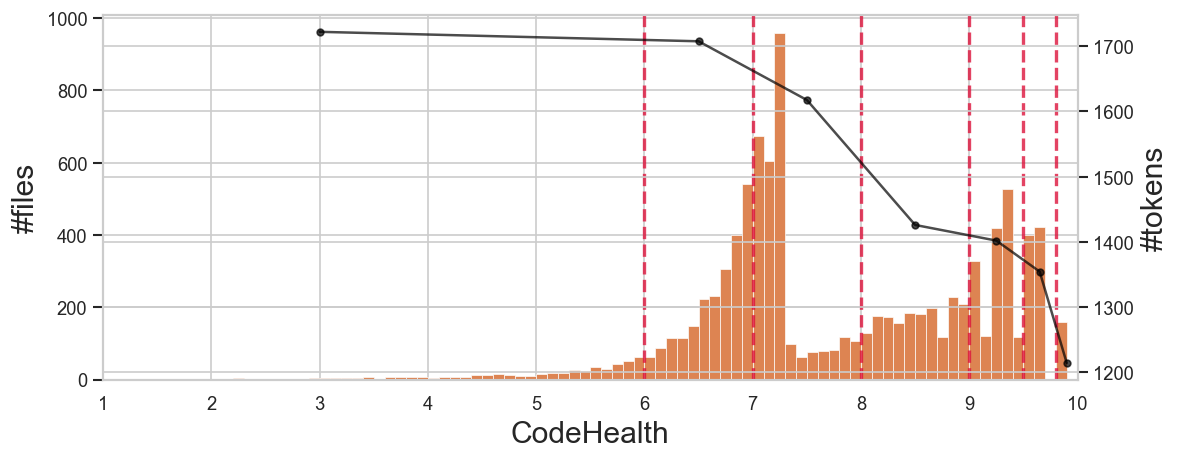}
        \caption{C++ (n=9,825)}
        \label{fig:ch-dist-cpp}
    \end{subfigure}
    \vspace{0.5em}
    \caption{Distribution of CH across the datasets. Dashed vertical lines show bin thresholds. The black line shows the median input tokens per bin (Claude Sonnet 4.6 tokenizer).}
    \label{fig:ch-distributions}
\end{figure}

Fig.~\ref{fig:ch-distributions} shows the CH distributions for Python, Java, and C++. Consistent with previous research~\cite{borgIndustrialCodeQuality2025}, the distributions are concentrated toward the higher end of the CH scale, with comparatively fewer files in the low-CH range. For Java and C++, our sampling strategy deliberately includes more lower-CH files, making the left-skewed shape more visible.

In the remainder of this paper, we discuss seven CH bins. The main threshold is $CH=9$, which denotes ``healthy'' code in CodeScene and has been calibrated over several years to align with human perceptions of maintainability~\cite{borgGhostEchoesRevealed2024a}. Our research~\cite{borgCodeMachinesNot2026} and experience with customers adopting AI-assisted development suggest that effective LLM-driven development may require even higher CH. Within healthy code ($CH \geq 9.0$), we further distinguish ordinary healthy code ($9 \leq CH < 9.5$), AI-ready code ($9.5 \leq CH < 9.8$), and near-perfect code ($CH \geq 9.8$). The remaining bins cover progressively lower ranges: 8--9, 7--8, 6--7, and $CH \leq 6$.

\subsection{LLM and Prompt Engineering}
We study test case generation using Qwen3-Coder-30B-A3B-Instruct (hereafter, Qwen3) an open-weight code LLM (temp=0.1, max tokens=4,096). The model uses a mixture-of-experts architecture with approximately 30.5B total parameters and 3.3B activated parameters per token. We selected Qwen3 because it showed promising results in our original study~\cite{borgCodeMachinesNot2026} and provided a practical tradeoff between capability and hardware requirements.

Previous research has shown that prompts can substantially influence the outcome of LLM-based coding tasks~\cite{fanLargeLanguageModels2023}, including processing of test code~\cite{gaoAutomatedUnitTest2025}. We therefore iteratively refined language-specific prompt templates for Python, Java, and C++. The goal was to increase the likelihood that Qwen3 generated syntactically valid test code in the expected format, so that the outputs could be executed by the corresponding unit test frameworks.

All prompts instructed the model to generate a single unit test file, output only raw source code, and avoid markdown fences or explanatory prose. They also specified the expected testing framework for each language: pytest for Python, JUnit~5 for Java, and CppUnit for C++. To reduce invalid outputs, the prompts included language-specific constraints, such as exact test class or module names, required imports, entry-point conventions, and handling of standard input/output. The source code under test was appended directly to the prompt. The exact prompt templates are available in the replication package~\cite{replication}.

\subsection{Experimental Setup} \label{sec:method_exp}
The experiments were executed on an NVIDIA A100 Tensor Core GPU with 80GB VRAM. We deployed Qwen3 locally using vLLM as the inference engine.

We implemented a batch-processing pipeline that iterates over all input files and generates unit tests in a single-shot setting. For each file, the pipeline prompts Qwen3 using the corresponding language-specific template and logs the prompt, raw model response, token usage, execution time, and success or failure status.

Despite careful prompt instructions, some model outputs contained formatting artifacts. We therefore applied deterministic post-processing before execution. This step removed markdown code fences and prose outside the generated code, validated that responses were non-empty, and wrote test files using deterministic naming schemes. No additional repair or correction was performed.

\subsection{Data Analysis}
We analyze the generated test cases using line coverage, branch coverage, and mutation score. For each outcome, we compute Spearman correlations with CH. Since file size is known to be inversely related to CH~\cite{tornhillCodeRedBusiness2022}, we also compute partial Spearman correlations controlling for Source Lines of Code (SLOC). We report significance as $^{*}p < .05$, $^{**}p < .01$, $^{***}p < .001$, and $ns$ for $p \geq .05$.

\subsubsection{Coverage Testing}
Coverage was measured separately for each target language using commonly used tooling: \texttt{coverage.py} with branch tracking for Python, \texttt{JaCoCo} with the \texttt{JUnit~5} console runner for Java, and \texttt{gcov}/\texttt{lcov} with \texttt{CppUnit} for C++. Each generated test file was compiled or executed together with its corresponding source file, and source-test pairs were isolated to avoid interference between files. Finally, we extracted line and branch coverage from the tools' resulting reports.
%For C++, source files defining their own \texttt{main} function were adjusted during compilation to avoid conflicts with the \texttt{CppUnit} test runner. 

%\subsubsection{Coverage Testing}
%For Java, we used \texttt{JaCoCo} together with the \texttt{JUnit~5} console runner. Each generated test file was compiled against its corresponding source file and executed with the \texttt{JaCoCo} Java agent enabled. To avoid class-name conflicts between files, each source-test pair was placed into an isolated package before compilation. The individual \texttt{JaCoCo} execution files were then merged into a combined report, from which line and branch coverage were extracted.

%For Python, we used \texttt{coverage.py} with branch tracking enabled. Generated pytest files were first checked for syntactic validity and then executed individually under coverage instrumentation. The resulting coverage data files were combined into a single coverage report.

%For C++, we used \texttt{gcov}/\texttt{lcov} together with \texttt{CppUnit}. Each generated test file was compiled with coverage instrumentation into a separate executable together with its corresponding source file. When a file defined its own \texttt{main} function, this function was renamed during compilation to avoid conflicts with the \texttt{CppUnit} test runner. Each executable was then run independently, and \texttt{lcov} was used to collect line and branch coverage.

\subsubsection{Mutation Testing}
Mutation scores were measured separately for each target language using language-specific tooling: \texttt{mutmut} for Python, \texttt{PIT} with the \texttt{JUnit~5} plugin for Java, and \texttt{Mull} with \texttt{CppUnit} for C++. For each source-test pair, we first executed the generated tests as a baseline and performed mutation testing only when the baseline tests passed. Java source-test pairs were evaluated in temporary \texttt{Maven} projects, Python pairs in temporary directories, and C++ pairs as temporary executables compiled with \texttt{g++} and the \texttt{Mull} compiler plugin. Across all languages, the mutation score was computed as the ratio of killed mutants to all generated mutants. Test suites that failed to compile, could not be executed, or timed out were recorded as unsuccessful executions and excluded from mutation-score computation.

\subsubsection{Tokenization}
To study input-token counts in relation to RQ\textsubscript{2}, we counted source-file tokens using four contemporary model families: Qwen, OpenAI, Claude, and Gemini. Specifically, we used the tokenizer distributed with Qwen, OpenAI's \texttt{o200k\_base} tokenizer, and the official token-counting APIs for Claude Sonnet~4.6 and Gemini~3.1~Pro Preview. Qwen and OpenAI counts were computed locally using Hugging Face Transformers and \texttt{tiktoken}, respectively. We then calculated Spearman correlations between CH and token counts, including partial correlations controlling for SLOC and character count.

\section{Results and Discussion} \label{sec:res}
This section analyzes line coverage, branch coverage, and mutation score for the generated test cases across the CH bins. Fig.~\ref{fig:effectiveness} shows the test effectiveness metrics across different CH bins, and Table~\ref{tab:combined_results} presents further details. Moreover, the section presents an analysis of token counts across tokenizers. 

\subsection{RQ1: CodeHealth and Test Effectiveness}
 %For each figure, we report Spearman's rank correlation coefficient together with its significance level.

\begin{figure}[t]
    \centering

    \begin{subfigure}[b]{\linewidth}
        \centering
        \includegraphics[width=\linewidth]{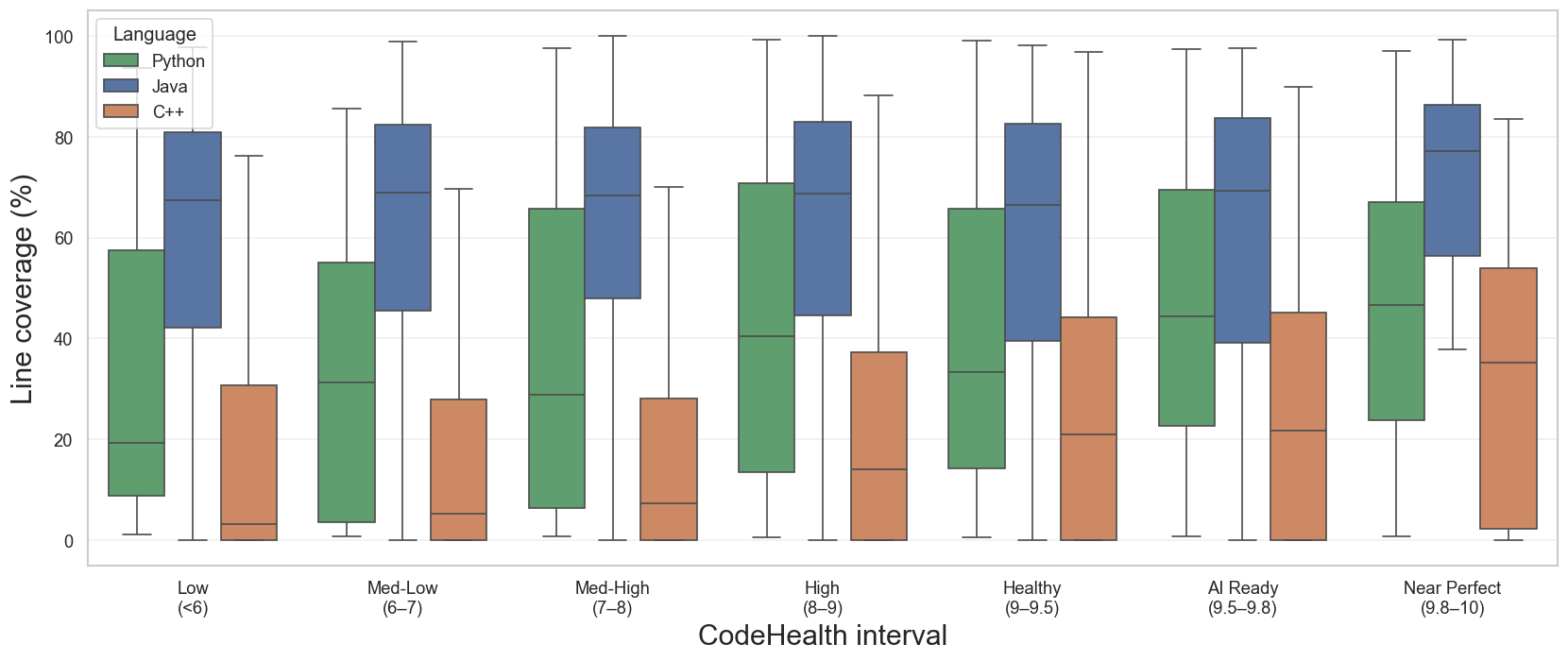}
        \caption{Line coverage}
        \label{fig:line-coverage}
    \end{subfigure}

    \vspace{0.5em}

    \begin{subfigure}[b]{\linewidth}
        \centering
        \includegraphics[width=\linewidth]{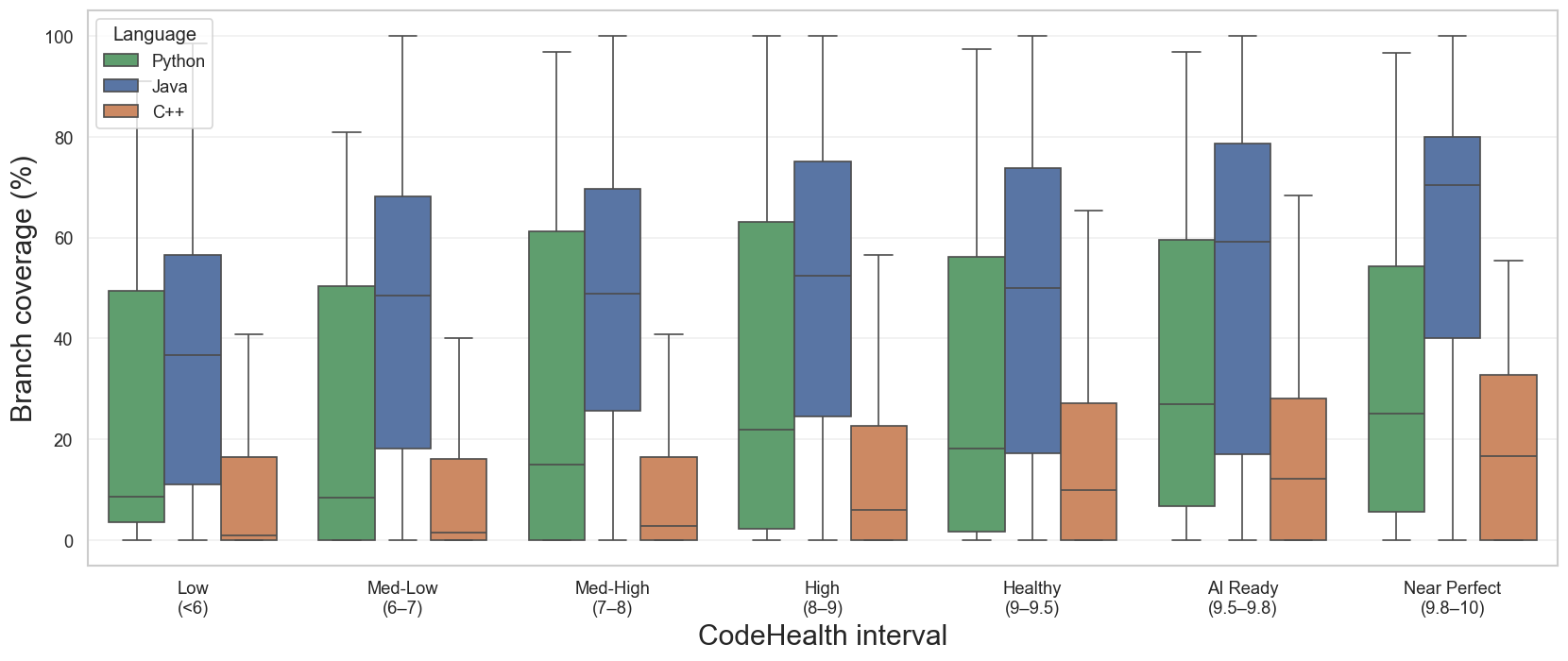}
        \caption{Branch coverage}
        \label{fig:branch-coverage}
    \end{subfigure}

    \vspace{0.5em}

    \begin{subfigure}[b]{\linewidth}
        \centering
        \includegraphics[width=\linewidth]{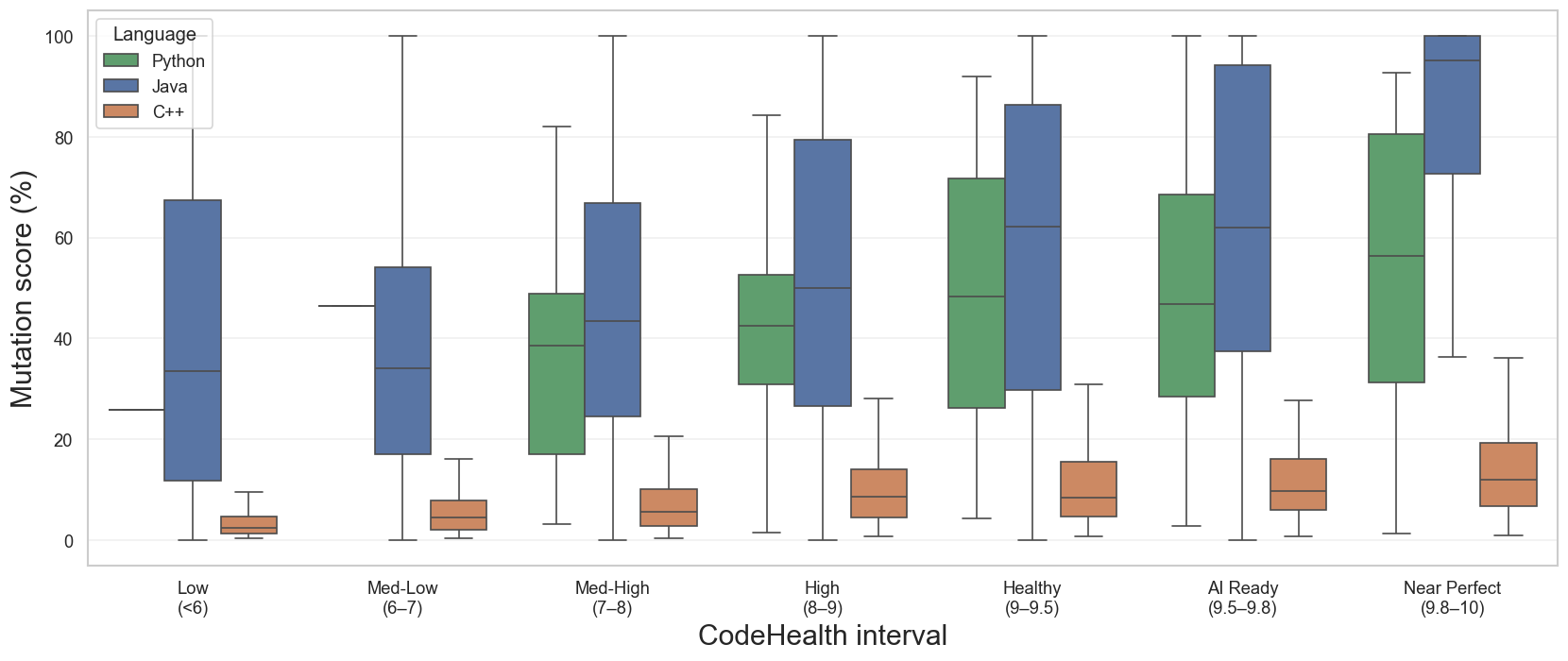}
        \caption{Mutation score}
        \label{fig:mutation}
    \end{subfigure}

    \caption{Test effectiveness metrics across different CH bins.}
    \label{fig:effectiveness}
\end{figure}

Fig.~\ref{fig:line-coverage} presents the line coverage achieved by the generated test cases. For Python, the median coverage scores are relatively stable across the CH intervals with at least 10 files, ranging from approximately 30\% to 47\% ($\rho = 0.11$). For Java, the medians remain consistently higher, between approximately 66\% and 77\% across the full CH range, with negligible correlation between CH and coverage ($\rho = 0.01$, ns). C++ shows substantially lower coverage, especially for files with low CH. However, it also exhibits the clearest increasing trend: the median values increase monotonically across CH intervals, although the correlation remains weak ($\rho = 0.15$), as presented in Table~\ref{tab:effect-correlations}.

\begin{table*}[t]
  \centering
  \caption{Test case generation results per CH bin, showing median line coverage~(\%), branch coverage~(\%), mutation score~(\%), and median input tokens. Values for Python in the CH$<$7 bins are very few and presented in italic font.}
  \label{tab:combined_results}
  \renewcommand{\arraystretch}{0.87}
  \setlength{\tabcolsep}{3pt}
  \setlength{\aboverulesep}{2pt}
  \setlength{\belowrulesep}{2pt}
  \footnotesize
  \begin{tabular}{@{} l |
                      r r r r r r |
                      r r r r r r |
                      r r r r r r @{}}
    \toprule
    & \multicolumn{6}{c}{\textbf{Python} ($n=5{,}000$)}
    & \multicolumn{6}{c}{\textbf{Java} ($n=9{,}795$)}
    & \multicolumn{6}{c}{\textbf{C++} ($n=9{,}825$)} \\
    \cmidrule(lr){2-7}\cmidrule(lr){8-13}\cmidrule(lr){14-19}
    \textbf{CH}
    & \textbf{$N$} & \textbf{SLOC} & \textbf{Line} & \textbf{Branch} & \textbf{Mut.} & \textbf{Tokens}
    & \textbf{$N$} & \textbf{SLOC} & \textbf{Line} & \textbf{Branch} & \textbf{Mut.} & \textbf{Tokens}
    & \textbf{$N$} & \textbf{SLOC} & \textbf{Line} & \textbf{Branch} & \textbf{Mut.} & \textbf{Tokens} \\
    \midrule
    $<$6
      &   \textit{10} &  \textit{87} & \textit{19.4} &  \textit{8.6} & \textit{25.9} & \textit{1{,}263.5}
      &   237 & 130 & 67.3 & 36.7 & 34.8 & 1{,}492.0
      &   479 & 123 &  3.2 &  0.9 &  2.5 & 1{,}598.0 \\
    6--7
      &   \textit{28} &  \textit{84} & \textit{31.3} &  \textit{8.3} & \textit{46.4} & \textit{1{,}072.0}
      &   893 & 149 & 68.9 & 48.5 & 34.4 & 1{,}426.0
      & 2{,}096 & 140 &  5.2 &  1.6 &  4.4 & 1{,}561.5 \\
    7--8
      &   277 &  89 & 28.9 & 15.0 & 38.6 & 1{,}056.0
      & 4{,}127 & 151 & 68.3 & 48.9 & 43.4 & 1{,}364.0
      & 2{,}990 & 136 &  7.3 &  2.9 &  5.7 & 1{,}486.0 \\
    8--9
      & 2{,}185 &  79 & 40.5 & 21.9 & 42.6 & 986.0
      & 2{,}038 & 142 & 68.7 & 52.4 & 50.0 & 1{,}252.0
      & 1{,}760 & 121 & 14.1 &  6.1 &  8.6 & 1{,}317.5 \\
    9--9.5
      &   994 &  78 & 33.3 & 18.2 & 48.3 & 985.0
      & 1{,}270 & 136 & 66.4 & 50.0 & 62.5 & 1{,}194.5
      & 1{,}518 & 120 & 20.9 & 10.0 &  8.5 & 1{,}290.0 \\
    9.5--9.8
      &   942 &  75 & 44.5 & 26.9 & 46.8 & 976.0
      &   959 & 134 & 69.4 & 59.1 & 62.5 & 1{,}171.0
      &   822 & 118 & 21.7 & 12.1 &  9.6 & 1{,}241.5 \\
    9.8--10
      &   564 &  74 & 46.6 & 25.0 & 57.8 & 933.0
      &   271 & 119 & 77.1 & 70.8 & 95.1 & 1{,}028.0
      &   160 & 115 & 35.2 & 16.7 & 12.0 & 1{,}111.5 \\
    \bottomrule
  \end{tabular}
\end{table*}

Fig.~\ref{fig:branch-coverage} presents the branch coverage achieved by the generated test cases. Compared with line coverage, branch coverage is generally lower, as expected. For Python, the median values remain modest across the CH intervals, ranging from 8\% to 27\%, with only a weak positive correlation ($\rho = 0.07$). Java again achieves the highest coverage levels. Unlike the line coverage results, however, the branch coverage medians show a clearer upward tendency across CH intervals, although the correlation remains very weak ($\rho = 0.08$). For C++, branch coverage remains low in absolute terms, but increases more consistently with CH than for Python and Java, yielding a weak positive correlation ($\rho = 0.14$).

\begin{table}[t]
\centering
\caption{CH-test effectiveness correlations. $\rho_{\mathrm{S}}$ is the partial correlation controlling for SLOC.}
\label{tab:effect-correlations}
\scriptsize
\setlength{\tabcolsep}{3pt}
\renewcommand{\arraystretch}{1.05}
\begin{tabular}{@{}lcccccc@{}}
\toprule
Language & \multicolumn{2}{c}{Line cov.} & \multicolumn{2}{c}{Branch cov.} & \multicolumn{2}{c}{Mutation score} \\
\cmidrule(lr){2-3}\cmidrule(lr){4-5}\cmidrule(l){6-7}
& $\rho$ & $\rho_{\mathrm{S}}$ & $\rho$ & $\rho_{\mathrm{S}}$ & $\rho$ & $\rho_{\mathrm{S}}$ \\
\midrule
Python & $+.11^{***}$ & $+.11^{***}$ & $+.07^{***}$ & $+.07^{***}$ & $+.18^{**}$ & $+.12$ \\
Java   & $+.01$       & $-.02^{*}$    & $+.08^{***}$ & $+.06^{***}$ & $+.21^{***}$ & $+.20^{***}$ \\
C++    & $+.15^{***}$ & $+.16^{***}$  & $+.14^{***}$ & $+.14^{***}$ & $+.34^{***}$ & $+.35^{***}$ \\
\bottomrule
\end{tabular}
\end{table}

Fig.~\ref{fig:mutation} shows the mutation scores achieved by the generated test cases. Mutation scores are only reported for files where the generated baseline tests passed, since mutation testing requires an executable test suite. 
This conditioning may bias the observed CH associations if execution success itself varies with CH, particularly in sparsely populated intervals. %As a result, some CH intervals, especially low-CH intervals with few successful test executions, contain few or no mutation-score observations. 
The mutation-score boxplots show an upward shift in central tendency, especially for Java and C++.

For Python, mutation testing generated results for only 251 files, and the median mutation scores range from approximately 39\% to 57\%, with a weak positive correlation ($\rho = 0.18$). For Java, the trend is more pronounced: the median mutation score increases from about 35\% in the lowest CH interval to about 95\% in the highest CH interval, resulting in a stronger correlation than for line and branch coverage ($\rho = 0.21$). For C++, the absolute mutation scores remain substantially lower, but they increase steadily across CH intervals, from approximately 3\% to 12\%. This yields the strongest observed association among the three languages ($\rho = 0.34$). 

Across the languages, the boxplots show substantial variability and overlapping distributions across CH intervals. Thus, the observed trends should not be interpreted as deterministic effects at the individual-file level. Instead, they indicate that higher CH is associated with a gradual shift in the distribution toward generating more effective test cases. This pattern is weak for coverage-based metrics, especially for Python and Java, but clearer for mutation score. Actually, within each language, $\rho$ tends to increase as the metric becomes more semantically demanding, i.e., from standard coverage metrics to mutation scores. This suggests that CH is more strongly related to behavioral test effectiveness than to coverage alone: mutation score captures whether tests detect injected behavioral changes, whereas coverage only captures whether code elements are executed.

Partial Spearman correlations controlling for SLOC ($\rho_{\text{S}}$) show that the main pattern remains largely stable. The coverage correlations are essentially unchanged for Python and C++, while the Java coverage associations weaken slightly after the control. For mutation testing, C++ remains the strongest observed association ($\rho_{\text{S}}=0.35$), and Java remains practically unchanged ($\rho_{\text{S}}=0.20$). The main exception is Python, where the mutation-score correlation is no longer statistically significant after controlling for SLOC; however, this result is based on only 251 observations.

\subsection{RQ2: CodeHealth and Token Efficiency}
The input-token columns in Table~\ref{tab:combined_results} report the median number of input tokens per CH interval during the test generation experiment. We find that in the lowest CH interval, the median input-token count is 45.1\% higher for Java and 43.8\% higher for C++ compared with the near-perfect interval. For Python, the pattern is less clear, which might partly be explained by the dataset containing very few files in the lowest CH intervals.

Table~\ref{tab:effic-correlations} shows detailed correlation analyses from contemporary tokenizers. Values with $|\rho| > .20$ are shown in bold. Furthermore, Fig.~\ref{fig:ch-distributions} depicts the median input tokens for the Claude Sonnet 4.6 tokenizer across the CH bins.

\begin{table}[t]
\centering
\caption{CH--token count correlations. $\rho_{\mathrm{S}}$ and $\rho_{\mathrm{C}}$ are partial correlations controlling for SLOC and character count, resp.}
\label{tab:effic-correlations}
\scriptsize
\setlength{\tabcolsep}{4pt}
\renewcommand{\arraystretch}{1.03}
\begin{tabular}{@{}lllll@{}}
\toprule
Lang. & Tokenizer & $\rho$ & $\rho_{\mathrm{S}}$ & $\rho_{\mathrm{C}}$ \\
\midrule
Python & Qwen        & $-.12^{***}$ & $.00$        & $-.02$    \\
       & OpenAI  & $-.12^{***}$ & $.00$        & $-.02$        \\
       & Claude    & $-.11^{***}$ & $+.01$       & $.00$         \\
       & Gemini    & $-.09^{***}$ & $+.04^{**}$  & $+.04^{**}$   \\
\addlinespace[1pt]
Java   & Qwen        & $\mathbf{-.30}^{***}$ & $\mathbf{-.34}^{***}$ & $\mathbf{-.28}^{***}$  \\
       & OpenAI  & $\mathbf{-.30}^{***}$ & $\mathbf{-.33}^{***}$ & $\mathbf{-.28}^{***}$  \\
       & Claude   & $\mathbf{-.29}^{***}$ & $\mathbf{-.32}^{***}$ & $\mathbf{-.27}^{***}$  \\
       & Gemini   & $\mathbf{-.28}^{***}$ & $\mathbf{-.30}^{***}$ & $\mathbf{-.25}^{***}$  \\
\addlinespace[1pt]
C++    & Qwen        & $\mathbf{-.28}^{***}$ & $\mathbf{-.28}^{***}$ & $-.13^{***}$  \\
       & OpenAI  & $\mathbf{-.28}^{***}$ & $\mathbf{-.29}^{***}$ & $-.14^{***}$  \\
       & Claude    & $\mathbf{-.25}^{***}$ & $\mathbf{-.25}^{***}$ & $-.05^{***}$  \\
       & Gemini   & $\mathbf{-.25}^{***}$ & $\mathbf{-.24}^{***}$ & $-.04^{***}$  \\
\bottomrule
\end{tabular}
\end{table}

Across all four tokenizers, lower CH is associated with higher input-token counts for Java and C++ ($\rho = -.25$ to $-.30$). The Python subset, which is concentrated in the higher CH intervals, shows only weak correlations ($\rho \approx -.1$). For these files, the association essentially disappears after controlling for SLOC. This suggests that lower-CH Python files require more tokens mainly because they contain more source lines.

For Java and C++, the negative correlations remain after controlling for SLOC ($\rho_{\mathrm{S}} = -.34$ to $-.24$). Thus, at a comparable number of SLOC, lower-CH Java and C++ files still tend to consume more tokens. The token overhead of lower CH is therefore not merely a line-count effect.

When controlling for character count ($\rho_{\mathrm{C}}$), the two languages differ. For Java, the association between CH and token count remains strong, i.e., at a comparable number of characters, lower-CH Java still tends to require more tokens. Low-CH Java appears to fragment into more tokens per character than high-CH Java. For C++, the character-controlled association weakens substantially, especially for the Claude and Gemini tokenizers. This suggests that, in competitive programming, lower-CH C++ files are more token-expensive mainly because they are more text-heavy. A complementary tokens-per-character analysis (replication package~\cite{replication}) supports this reading: CH correlates negatively with tokens per character for Java across all four tokenizers, but for C++ only weakly and only for the Qwen and OpenAI tokenizers.

We hypothesize that Java syntax and code conventions might influence the difference in ($\rho_{\mathrm{C}}$). Java code often relies on chained method calls~\cite{keshkMethodChainingRedux2023} and generic types~\cite{parninAdoptionUseJava2013}, both of which may contribute to token overhead. Moreover, Java's camelCase identifiers~\cite{zhangEmpiricalStudyUsage2021} tend to fragment into several tokens. On the other hand, C++ has compact syntactic mechanisms whose complexity may not expand input tokens as directly as Java idioms do. Examples include pointer use, macros, and operator-heavy expressions. Future work is clearly needed to investigate this in more detail.

%\begin{RQBoxHeader}{RQ$_2$: How are token counts related to CodeHealth?}
%CH is negatively correlated with input-token count across tokenizers. For Java and C++, the association remains after controlling for SLOC. For Java, it also remains after controlling for character count, suggesting that low-CH Java code is less token-efficient per character.
%\end{RQBoxHeader}

\section{Implications and Limitations} \label{sec:impl}
Coding agents' ability to analyze and manipulate source code in practical settings does not depend only on the capabilities of their underlying LLMs, agent harnesses, and instructions. The maintainability of the code they operate on is also a critical factor. In our previous work~\cite{borgCodeMachinesNot2026}, we used the success rate of AI-generated refactoring as a proxy for AI-friendliness~\cite{thoughtworksTechnologyRadarOpinionated2025}. In this study, we extend that perspective to test-case generation by analyzing whether healthier code is associated with more effective generated tests.

Our results related to RQ\textsubscript{1} add more evidence to the proposition that AI tools, just like humans, perform better in healthy codebases. Median line coverage, branch coverage, and mutation score are highest in the upper CH bins, and the correlation between CH and mutation score in particular suggests a potentially relevant implication for Java and C++.

For RQ\textsubscript{2}, our findings provide an initial link between software maintainability and LLM economics. Poor maintainability is traditionally understood as a human cost: it makes code harder to understand, test, and evolve. In line with our previous study~\cite{borgCodeMachinesNot2026}, our analysis suggests an additional AI-era cost: unhealthy code may also consume more LLM context.

This matters for AI-assisted development because tokens are the unit on which LLM APIs are billed, affecting prompting costs, code retrieval, context-window usage, and the practical efficiency of agentic workflows. Poor maintainability may therefore create a double penalty, i.e., it challenges human cognition while also making LLM-driven work more costly.

This study constitutes a first exploration of test effectiveness and token efficiency. The main threat to the validity of this work is external validity, as the source code under study originates in competitive programming. CodeContests consists of standalone competitive-programming solutions rather than conventional production units, limiting generalizability to industrial unit testing. 
%Coding under such circumstances exhibits a distinct setting: it's all about providing a working solution under extreme time pressure. 
However, the corpus spans many quality levels, and CH captures local properties (nesting, complexity, etc.) that exist in any code regardless of domain. Also, our effectiveness scores are comparable to those reported in related work~\cite{zhangLargeLanguageModels2025,jainTestGenEvalRealWorld2025}, thus we argue that the relative differences across bins carry meaning.

A second threat to generalizability is that we use a single LLM, Qwen3-Coder-30B. Larger frontier models might generate substantially better test cases, potentially also for lower-CH code. Given the rapid development of AI-assisted coding, future work should study code from additional domains, newer LLMs, and more advanced agentic workflows.

\section{Conclusion and Future Work} \label{sec:conc}
First, CH provides a \textit{weak but consistent} signal of LLM-generated test effectiveness. The relationship is clearest for mutation score, suggesting that healthier code is more likely to yield test cases that detect injected behavioral changes, not only tests that execute lines or branches. Second, CH is negatively correlated with input-token count across tokenizers. For Java and C++, the association remains after controlling for SLOC. For Java, it also remains after controlling for character count, suggesting that low-CH Java code incurs token overhead beyond source-text length alone.

This preliminary work opens avenues for future research. First, we plan to expand the set of studied LLMs and move beyond the domain of competitive programming. The current setup was useful for identifying differences in how LLMs process code across varying levels of CH. However, future studies should consider more realistic testing workflows based on agentic designs, including test-suite execution and iterative test generation toward coverage targets. Such multi-turn settings would also provide a richer context for studying token economics, including output-token consumption, which was considered out of scope in the present study.

\section*{Acknowledgment}
This work was partially supported by the Wallenberg AI, Autonomous Systems and Software Program (WASP WARA-Ops) and the Competence Centre NextG2Com (VINNOVA grant 2023-00541).

%This work was partially supported by the Wallenberg AI, Autonomous Systems and Software Program WASP (WARA-Ops), funded by the Knut and Alice Wallenberg Foundation, and partly by the Competence Centre NextG2Com funded by the VINNOVA program for Advanced Digitalisation with grant number 2023-00541.


\newpage
\bibliographystyle{IEEEtran}
\bibliography{testgen}

\end{document}